\documentstyle[12pt,epsf]{article}

\begin{document}

\baselineskip=18pt plus 0.2pt minus 0.1pt

\makeatletter

\def\@bgnmark{<}
\def\@endmark{>}
\def\WKht{.85}
\def\WKsep{.4}
\def\WKrule{.03}
\newcount\@bgncnt
\newcount\@endcnt
\newcount\@h@ight
\newcount\TempCount
\newif\if@Exist
\newdimen\@tempdimc
\newdimen\@tempdimd
\newdimen\h@ight 
\newdimen\w@dth  
\def\sqrt{\radical"270370}

\def\SEPbgn#1<#2#3#4\@@{\xdef\@MAE{#1}\xdef\@MARK{#2}
\xdef\@FRONT{#3}\xdef\@USIRO{#4}}
\def\SEPend#1>#2#3#4\@@{\xdef\@MAE{#1}\xdef\@MARK{#2}
\xdef\@FRONT{#3}\xdef\@USIRO{#4}}
\def\c@lc{
 \setbox0=\hbox{$\displaystyle \@FRONT$}
 \@tempdima\wd0 \@tempdimb\ht0
 \settowidth{\@tempdimc}{$\displaystyle \@MAE$}
 \settowidth{\@tempdimd}{$\displaystyle \@list$}
 \divide\@tempdima by2 \advance\@tempdima by \@tempdimc
 \advance\@tempdima by \@tempdimd}

\def\@dblfornoop#1\@@#2#3#4{}
\def\@dblfor#1;#2:=#3\do#4{\xdef\@fortmp{#3}\ifx\@fortmp\@empty \else%
 \expandafter\@dblforloop#3\@nil,\@nil,\@nil\@@#1#2{#4}\fi}
\def\@dblforloop#1,#2,#3\@@#4#5#6{\def#4{#1} \def#5{#2}%
 \ifx #4\@nnil \let\@nextwhile=\@dblfornoop \else%
 #6\relax \let\@nextwhile=\@dblforloop\fi\@nextwhile#3\@@#4#5{#6}}

\def\fin@endpt#1#2{
\@dblfor\MemBer;\NextmemBer:=#2\do{\def\@bject{#1}%
 \if \MemBer\@bject \xdef\@endpt{\NextmemBer} \@Existtrue\fi}}%
\def\fin@h@ight#1#2{
 \@tempcnta\z@%
 \@tfor\MEmber:=#2\do{\advance\@tempcnta\@ne%
 \ifnum \@tempcnta=#1 \@h@ight=\MEmber\fi}}

\def\wicksymbol#1#2#3#4#5{
 \@tempdima=#3 \advance\@tempdima-#1%
 \@tempdimc=#5\h@ight \@tempdimb=\@tempdimc \advance\@tempdimb-\w@dth%
 \@tempdimd=#2 \advance\@tempdimd-1.587ex
 \hskip#1%
 \vrule height \@tempdimc width\w@dth depth-\@tempdimd \kern-\w@dth%
 \vrule height \@tempdimc width\@tempdima depth-\@tempdimb\kern-\w@dth%
 \@tempdimd=#4 \advance\@tempdimd-1.587ex 
 \vrule height \@tempdimc width\w@dth depth-\@tempdimd}

\def\first#1{\expandafter\@mae#1\@nil}
\def\secnd#1{\expandafter\@ato#1\@nil}
\def\@mae#1;#2\@nil{#1}
\def\@ato#1;#2\@nil{#2}

\def\wick#1#2{%
 \h@ight=\WKht ex \w@dth=\WKrule em%
 \def\@wickdata{} \def\bgnend@list{} \@bgncnt\z@ \@endcnt\z@%
 \def\@list{} \def\bgnp@sition{} \def\endp@sition{}%
 \xdef\str@ng{#2}
 \@tfor\m@mber:=#2\do{%
 \ifx\m@mber\@bgnmark \advance\@bgncnt\@ne
  \expandafter\SEPbgn\str@ng\empty\@@ \c@lc 
  \xdef\bgnp@sition{\bgnp@sition\@MARK,\the\@tempdima;\the\@tempdimb,}
  \xdef\@list{\@list\@MAE\@FRONT}
  \xdef\str@ng{\@USIRO}\fi
 \ifx \m@mber\@endmark \advance\@endcnt\@ne
  \expandafter\SEPend\str@ng\empty\@@ \c@lc
  \xdef\endp@sition{\endp@sition\@MARK,\the\@tempdima;\the\@tempdimb,}
  \xdef\@list{\@list\@MAE\@FRONT}
  \xdef\str@ng{\@USIRO}\fi}
  \xdef\@list{\@list\@USIRO}
 \ifnum\@bgncnt=\@endcnt \else%
 \@latexerr{The numbers of `<' and `>' do not match}%
 {You have written different numbers of < and >}\fi%
 \TempCount\z@ \@tfor\mmbr:=#1\do{\advance\TempCount\@ne}%
 \ifnum\@bgncnt=\TempCount \else%
 \@latexerr{The number of numbers in the first argument is different
 with that of contractions <...>}%
 {Give the same numbers of heights as the contractions <...>}\fi
 \mathop{\vbox{\m@th\ialign{##\crcr\noalign{\kern\WKsep ex}%
 $\m@th \TempCount\z@%
 \@dblfor\member;\nextmember:=\bgnp@sition\do{
 \advance\TempCount\@ne \xdef\@bgnpt{\nextmember}%
 \@Existfalse%
 \fin@endpt{\member}{\endp@sition}%
 \if@Exist \else \@latexerr{The begin-mark `<\member' has no
corresponding end-mark `>\member'}{You should write coinciding label 
like <\member .. >\member}\fi%
 \fin@h@ight{\TempCount}{#1}%
 \setbox0=\hbox{%
 $\wicksymbol{\first\@bgnpt}{\secnd\@bgnpt}{\first\@endpt}%
 {\secnd\@endpt}{\@h@ight}$\hss}
 \dp0\z@ \wd0\z@ \box0%
 }$\crcr\noalign{\kern\WKsep ex\nointerlineskip}%
 \setbox0=\hbox{$\displaystyle\@list$}\ht0=1.587ex%
 \box0\crcr}}}\limits}

\@addtoreset{equation}{section}
\renewcommand{\theequation}{\thesection.\arabic{equation}}

\newcommand{\nn}{\nonumber}
\newcommand{\bF}{\ket{B(F)}}
\newcommand{\bFb}{\ket{B^{(b)}(F)}}
\newcommand{\BF}{B(F)}
\newcommand{\bFx}{\ket{B\Bigl(F(x)\Bigr)}}
\newcommand{\bFxb}{\ket{B^{(b)}\Bigl(F(x)\Bigr)}}
\newcommand{\BFx}{B\Bigl(F(x)\Bigr)}
\newcommand{\bnew}{\ket{{\cal B}}}
\newcommand{\Bnew}{{\cal B}}
\newcommand{\bnewA}{\ket{{\cal B}\Bigl(A_\mu(x)\Bigr)}}
\newcommand{\BnewA}{{\cal B}\Bigl(A_\mu(x)\Bigr)}
\newcommand{\bnewb}{\ket{{\cal B}^{(b)}}}
\newcommand{\Bnewb}{{\cal B}^{(b)}}
\newcommand{\bnewAb}{\ket{{\cal B}^{(b)}\Bigl(A_\mu(x)\Bigr)}}
\newcommand{\BnewAb}{{\cal B}^{(b)}\Bigl(A_\mu(x)\Bigr)}
\newcommand{\bnewf}{\ket{{\cal B}^{(f)}}}
\newcommand{\Bnewf}{{\cal B}^{(f)}}
\newcommand{\bnewAf}{\ket{{\cal B}^{(f)}\Bigl(A_\mu(x)\Bigr)}}
\newcommand{\BnewAf}{{\cal B}^{(f)}\Bigl(A_\mu(x)\Bigr)}
\newcommand{\Lz}{\Lambda\Bigl(\zeta_\mu(x)\Bigr)}
\renewcommand{\star}{*}
\newcommand{\tr}{\mathop{\rm tr}}
\newcommand{\QB}{Q_{\rm B}}
\newcommand{\Half}{\frac{1}{2}}
\newcommand{\bra}[1]{\left\langle #1\right|}
\newcommand{\hs}[1]{\hspace*{#1}}
\newcommand{\vs}[1]{\vspace*{#1}}
\newcommand{\ket}[1]{\left| #1\right\rangle}
\newcommand{\VEV}[1]{\left\langle #1\right\rangle}
\newcommand{\braket}[2]{\VEV{#1 | #2}}
\newcommand{\ac}{\overline{c}}
\newcommand{\calO}{{\cal O}}
\newcommand{\p}{\partial}
\newcommand{\wt}[1]{\widetilde{#1}}
\newcommand{\wtX}{\wt{X}}

\makeatother

\begin{titlepage}
\title{
\hfill\parbox{4cm}
{\normalsize KUNS-1599\\{\tt hep-th/9909095}}\\
\vspace{1cm}
Generalized Supersymetric Boundary State
\\[50pt]}
\author{
{\sc Koji Hashimoto}\thanks{{\tt hasshan@gauge.scphys.kyoto-u.ac.jp}}
\\[7pt]
{\it Department of Physics, Kyoto University, Kyoto 606-8502, Japan} 
}
\date{\normalsize September, 1999}
\maketitle
\thispagestyle{empty}

\begin{abstract}
\normalsize

Following our previous paper (hep-th/9909027), we generalize a
supersymmetric boundary state so that arbitrary
configuration of the gauge field coupled to the boundary of the
worldsheet is incorpolated. This generalized boundary state is BRST
invariant and satisfies the non-linear boundary conditions with
non-constant gauge field strength. This boundary state contains
divergence which is identical with the loop divergence in a
superstring $\sigma$ model. Hence vanishing of the $\beta$
function in the superstring $\sigma$ model corresponds to a
well-defined boundary state with no divergence. 
The coupling of a single closed superstring massless mode with
multiple open string massless modes is encoded in the boundary state,
and we confirm that derivative correction to the D-brane action in
this sector vanishes up to the first non-trivial order
$O(\alpha'\p^2)$. Combining T-dualities, we incorpolate also general 
configurations of the scalar fields on the D-brane, and construct
boundary states representing branes stuck to another D-brane, with use
of BIon configuration.  

\end{abstract}

\end{titlepage}

\section{Introduction}

Rrecent developmernt on D-branes in string theories is owing in some
part to the analysis of the boundary state
\cite{293,CLNY2,CLNY3,PC,Ishi,ML}.
This ingredient is effective to describe boundary of string
worldsheet, especially when it is studied in the ordinary
parturbative approach of closed string conformal field theory. The
boundary state itself has been developped to formulate string loop
corrections to the effective equations of motion of string theory in
relation to the Fischler-Susskind mechanism \cite{293,CLNY2,CLNY3}. In 
most of the studies of the boundary state, only the constant
gauge field strength was taken into account as the boundary degree of
freedom. This is because in that case  the boundary conditions
satisfied by the coordinate fields on the worldsheet become linear,
hence one can construct explicitly the eigen state of the equations of
the boundary conditions. 

The D-brane \cite{Pol} opened new directions 
on the use of the boundary state. Since this state specifies the
couplings between D-brane dynamical degrees of freedom and closed
string excitations, the state is useful for deriving D-brane actions,
the effective actions of the D-branes \cite{fdp,GG}. The D-brane
actions have been indicative of various dualities in string theories 
\cite{DD}. However, as mentioned above, these D-brane actions have
been obtained in only the leading order estimation with respect to the 
derivatives acting on the fields on the D-branes. The
derivative corrections may ruin the intriguing properties which
the D-brane actions possess, and hence a systematic method for
evaluating higher order corrections of the D-brane actions and the
boundary states are needed \cite{BBG,Tse,Kal}. 

In this paper, we extend our previous result of ref.\ \cite{KH} to the 
superstring case. We construct in superstring theory a 
generalized boundary state which incorporates whole degrees of freedom
of the boundary-coupled background fields, which are the gauge fields
and the scalar fields on the D-brane. This boundary state is BRST
invariant and reduces to the previously known form when the field
strength is put constant. Using the boundary state, we can calculate
particular terms in the D-brane action: the terms linear in closed
string massless modes. By generalizing the boundary state, we obtain
corrections to the relevant part of the D-brane action. This
method may give a gleam of hope to understand the string dualities to
all order.  

The organization of this paper is as follows.
In sec.\ \ref{sec:gene}, we obtain a supersymmetric extension of the
generalized boundary state given in our previous paper \cite{KH} in
which in the bosonic string theory a boundary gauge field (in
particular, non-constant modes of the field strength) was incorporated
to the boundary state. We treat type II superstring theory in this
paper. After checking the BRST invariance of the generalized boundary
state, in sec.\ \ref{Div} we evaluate the divergence immanent in this
state. The definition of the generalized boundary state includes
products of the coordinate scalar field on a single worldsheet point,
thus contains the normal-ordering divergence. In the evaluation of the
divergence, we adopt the approximation of slowly-varying fields. 
This divergence is found to be identical to the
superstring $\sigma$ model loop divergence, hence the vanishing of the 
$\beta$ function in the string $\sigma$ model corresponds to a
well-defined boundary state with no divergence. Then with use of this
well-defined state we find that there is no finite correction to the
D-brane action upto  this order $O(\alpha'\p^2)$ in the sector of the
coupling  linear in massless colsed string mode. This result is in
contrast to the bosonic string case \cite{KH}, in which the finite
corrections exist. In sec.\ \ref{sec:higgs}, utilizing T-duality
transformation we introduce scalar fields into the boundary state, and
see that there is no derivative correction to the D-brane action even
with the scalar fields. In appendix A, we present detailed
calculations which are necessary in sec.\ \ref{Div}. In appendix B, we
apply our generalized boundary state to the BIon configuration. We see
that the ``spike'' part of the BIon actually describes the stirng
stuck to the D-brane.

\section{Generalized supersymmetric boundary state}
\label{sec:gene}

As mentioned in the introduction, most aspects of the D-branes are
concerning supersymmetry and their BPS properties. Hence the
low energy effective actions of the D-branes are desirably studied in
a supersymmetric fashion. In this section, we apply the 
idea developped in our previous papar \cite{KH} to superstring theory
and construct a generalized supersymmetric boundary state,
incorporating the non-constant gauge field strength on the
D-brane. The scalar field which also exists on the D-brane shall be
treated later in sec.\ \ref{sec:higgs}. 


\subsection{Review of the generalized bosonic boundary state}

First let us summarize the definition and the relevant properties of
the generalized boundary state with the non-constant field strength in
bosonic string theory \cite{KH}. In the case of constant gauge field
strength $F_{\mu\nu}$, the boundary state $\bFb$ is defined as an
eigen state of the  linear boundary conditions for open bosonic
strings \cite{293,CLNY2,PC}:
\begin{eqnarray}
&&X^i(\sigma)\bFb=0 ,
\label{X_B=0}\\
&&\left(\pi P_\mu(\sigma)+F_{\mu\nu}\p_\sigma
X^\nu(\sigma)\right)\bFb=0 ,
\label{(P-FX)B}\\
&&\pi_c(\sigma)\bFb=\pi_{\ac}(\sigma)\bFb=0 .
\label{pi_c_B}
\end{eqnarray}
The index $i$ specifies the directions transverse to the D-brane,
while the Greek indices $\mu,\nu$ run in the longitudinal directions.
The superscript $(b)$ denotes that this is for the bosonic part.
The BRST invariance of this boundary state is a consequence of eqs.\
(\ref{X_B=0}) --- (\ref{pi_c_B}).
The oscillator representation of $\bFb$ reads 
\begin{eqnarray}
\bFb= 
-{T_p\over 4}N(F)
\ket{B_{\rm N}(F)}\otimes\ket{B_{\rm D}}\otimes\ket{B_{\rm gh}}
\label{bou}
\end{eqnarray}
where three factors of kets $\ket{B_{\rm N}(F)}$, $\ket{B_{\rm D}}$
and $\ket{B_{\rm gh}}$ are for satisfying (\ref{X_B=0}),
(\ref{(P-FX)B}) and (\ref{pi_c_B}) respectively. These are given by 
\begin{eqnarray}
&&\ket{B_{\rm N}(F)}=\exp \biggl\{ 
-\sum_{n\geq 1} {1\over n}
\alpha_{-n}^{(-)\mu}\calO (F)_\mu^{\;\; \nu}\alpha_{-n\;\nu}^{(+)}
\biggr\}\ket{0}_{p+1},
\label{aoa}\\
&&\ket{B_{\rm D}}=\exp \biggl\{ 
\sum_{n\geq 1} {1\over n}
\alpha_{-n}^{(-)i}\alpha_{-n\; i}^{(+)}
\biggr\}\ket{0}_{d-p-1}\delta^{d-p-1}(x^i),
\label{delta}\\
&&\ket{B_{\rm gh}}=\exp \biggl\{ 
\sum_{n\geq 1} 
(c_{-n}^{(-)}\bar{c}_{-n}^{(+)}+c_{-n}^{(+)}\bar{c}_{-n}^{(-)})
\biggr\}\ket{0}_{\rm gh}.
\label{ghostF}
\end{eqnarray}
In eq.\ (\ref{aoa}), The orthogonal matrix $\calO$ is defined as
$\calO_\mu^{\;\;\nu}= (1-F)_\mu^{\;\;\rho}
\{(1+F)^{-1}\}_{\rho}^{\;\;\nu}$.
The front factor $N(F)$ is obtained in various ways
\cite{CLNY2,CLNY3,our,front} as 
\begin{eqnarray}
\label{nf}
N(F)=\Bigl(\det (1+F)\Bigr)^{-\zeta(0)}.
\end{eqnarray}
When the front factor is put in this form (\ref{nf}), this boundary
state can be rewritten in a form in which the dependence on the gauge
field is simply arranged :
\begin{eqnarray}
  \bFb=\exp \left({i\over 2\pi}\oint\! d\sigma 
                  \, \p_\sigma X^\mu F_{\mu\nu}X^\nu\right)
\:\ket{B^{(b)}(F\!=\!0)}.
\label{arrange}
\end{eqnarray}

In ref.\ \cite{KH}, we have generalized this boundary state to the one 
with general configuration of the gauge field, using the properties of 
the star product (three-string interaction vertex) in a string field
theory. The definition of this generalized boundary state is 
\begin{eqnarray}
\label{Bnew}
\bnewAb\equiv \:U[A]\:\ket{B^{(b)}(F\!=\!0)},
\end{eqnarray}
with 
\begin{eqnarray}
\label{U[A]}
  U[A]\equiv \exp \left({-i\over \pi}\oint\! d\sigma 
                  \, \p_\sigma X^\mu A_{\mu}(X)
  \right).
\end{eqnarray}
One observes that, for the constant field strength, the generalized
boundary state (\ref{Bnew}) is reduced to the previous one
(\ref{arrange}). This boundary (\ref{Bnew}) state is subject to the
following non-linear boundary condition in the longitudinal directions
\begin{eqnarray}
\label{eigen}
\biggl[
\pi P_\mu + F_{\mu\nu}(X)
\p_\sigma X^\nu \biggr] \bnewb = 0,
\end{eqnarray}
in addition to the usual Dirichlet boundary conditions
$X^i \bnewb=0$. Note that in eq.\ (\ref{eigen}), the field
strength can be general, not only constant. Furthermore, we can show
that this $\ket{\cal B}$ is BRST invariant, 
\begin{eqnarray}
  Q_{\rm B}\bnewb=0,
\end{eqnarray}
with use of the boundary conditions above.


\subsection{Supersymmetric extension}


In the Neveu-Schwarz-Ramond (NS-R) formulation of superstring theory,
worldsheet fermions and superghosts are introduced in addition to the
contents in bosonic string theory. In this paper we shall not mention
the superghost part in the boundary state explicitly,  since this is
not concerned with the gauge field configuration. 

In the absence of non-trivial gauge field on the boundary, the
fermionic boundary conditions for the superstring worldsheet are
$(\psi_+ \mp i \psi_-)^\mu =0$ in the directions tangential to the
D-brane, and  $(\psi_+ \pm i \psi_-)^i =0$ for the transverse
directions \cite{293,CLNY2,PC}. The sign $\pm$ is the spin structure,
and we leave it undetermined.\footnote{Our result on the correction to
  the D-brane action does not refer to the spin structure as well as 
  other fine structures such as GSO projections.} Therefore the
fermionic sector of the 
boundary state $\bnewAf$ should satisfy
\begin{eqnarray}
  (\psi_+ \mp i \psi_-)^\mu\ket{{\cal B}^{(f)}(A\!=\!0)}=
  (\psi_+ \pm i \psi_-)^i \ket{{\cal B}^{(f)}(A\!=\!0)}=0.
\end{eqnarray}
The explicit oscillator representation of this eigen state is 
\begin{eqnarray}
\ket{{\cal B}^{(f)}(A\!=\!0)} = 
\exp 
  \left\{
    \mp i \sum_{r>0}b_{-r}^{(-)\mu}b_{-r\;\;\mu}^{(+)}
  \right\}
\exp 
  \left\{
    \pm i \sum_{r>0}b_{-r}^{(-)i}b_{-r\;\;i}^{(+)}
  \right\}
\ket{B^{(f)}_0} \otimes \ket{B_{\rm superghost}},
\end{eqnarray}
where $\ket{B^{(f)}_0}$ is the fermionic zero mode part which exists
only in the R-R sector. It is given as 
\begin{eqnarray}
  \ket{B^{(f)}_0} \equiv \Pi_i 
\theta^i\ket{0;\pm},
\label{fzero}
\end{eqnarray}
where we have defined a linear combination of the fermionic zero modes
$\theta^M\equiv\psi^M_{0+}\pm i\psi^M_{0-}$. The superghost sector 
$\ket{B_{\rm superghost}}$ is given in ref.\ \cite{293}.

Now, let us define the fermionic part of the generalized boundary
state as 
\begin{eqnarray}
  \bnewAf \equiv U[F] \ket{{\cal B}^{(f)}(A\!=\!0)}
\end{eqnarray}
where the unitary operator $U[F] \equiv \exp R$ is given by
\begin{eqnarray}
\label{Rdef}
R\equiv \frac{1}{8\pi} \oint d\sigma  
(\psi_+ \pm i \psi_-)^\mu (\psi_+ \pm i \psi_-)^\nu F_{\mu\nu}[X].
\end{eqnarray}
From the definition, this boundary state incorporates general
configuration of the gauge field strength $F_{\mu\nu}[X]$ and
evidently gauge invariant. Since the operator $U[F]$ is subject to the
following relation 
\begin{eqnarray}
  U[F] (\psi_+ \mp i \psi_-)^\mu  U[F]^{-1}
= (\psi_+ \mp i \psi_-)^\mu - F^\mu_{\;\;\nu}[X]
(\psi_+ \pm i \psi_-)^\nu,
\end{eqnarray}
then for the general background we obtain the non-linear boundary
condition for the tangential directions as
\begin{eqnarray}
\label{p-Fp}
 \biggl[ (\psi_+ \mp i \psi_-)^\mu - F^\mu_{\;\;\nu}[X]
(\psi_+ \pm i \psi_-)^\nu \biggr]
\bnewAf =0.
\end{eqnarray}
This boundary condition is actually the one given in ref.\
\cite{BSPT}, from the supersymmetrization of the boundary coupling in
the string $\sigma$ model.


Combining this fermionic part with the bosonic part in the previous
subsection, the total unitary operator becomes $U\equiv U[A] U[F]$. 
Since we have introduced general $X$ dependence in the fermionic
unitary operator $U[F]$, the boundary condition for the bosonic part
(\ref{eigen}) is changed. Actually, using the relation
\begin{eqnarray}
  [\pi P_\rho , R] =
\frac{-i}{8} 
(\psi_+ \pm i \psi_-)^\mu (\psi_+ \pm i \psi_-)^\nu 
\frac{\delta}{\delta X^\rho}F_{\mu\nu}[X],
\end{eqnarray}
we obtain
\begin{eqnarray}
\label{AFFA}
\lefteqn{  U[A] U[F] (\pi P_\mu) U[F]^{-1} U[A]^{-1}} \nn\\
&&=\pi P_\mu + F_{\mu\nu}[X] \p_\sigma X^\nu +\frac{i}{8} 
\p_\mu F_{\nu\rho}[X] (\psi_+ \pm i \psi_-)^\nu (\psi_+ \pm i
\psi_-)^\rho,
\end{eqnarray}
thus the boundary condition is modified by the third term on
the right hand side (RHS). We summarize the boundary conditions
as 
\begin{eqnarray}
&&  \biggl[\pi P_\mu + F_{\mu\nu}[X] \p_\sigma X^\nu+\frac{i}{8} 
\p_\mu F_{\nu\rho}[X] (\psi_+ \pm i \psi_-)^\nu (\psi_+ \pm i
\psi_-)^\rho\biggr]
\bnew=0,\label{p+FXB}\\
&& \biggl[ (\psi_+ \mp i \psi_-)^\mu - F^\mu_{\;\;\nu}[X]
(\psi_+ \pm i \psi_-)^\nu \biggr]
\bnew =0, \label{p-FpB}
\end{eqnarray}
where we have defined the complete boundary state as 
\begin{eqnarray}
\bnew \equiv \bnewb \otimes \bnewf.
\label{defwhole}
\end{eqnarray}

Of course when the field strength is constant, the generalized
boundary state (\ref{defwhole}) reproduces the usually utilized
boundary state. For the bosonic sector of the boundary state, this is
checked in ref.\ \cite{KH}. For the fermionic part, exponentiating the
linear operator $R$ (\ref{Rdef}), we have an explicit form of the
boundary state which is written only by the creation operators as
\begin{eqnarray}
  \ket{B^{(f)}(F)} = N^{(f)}(F) \exp 
  \left\{
    \mp i \sum_{r>0}b_{-r}^{(-)\mu}
        \calO_\mu^{\;\nu} b_{-r\;\;\nu}^{(+)}
  \right\}\ket{B^{(f)}_0(F)}\otimes \ket{B_{\rm superghost}}.
\label{expo}
\end{eqnarray}
In the R-R sector, the fermionic zero mode is modified and depends on
the constant field strength as 
\begin{eqnarray}
  \ket{B^{(f)}_0(F)} \equiv \exp
  \left(
    \frac14 \theta^\mu \theta^\nu F_{\mu\nu} 
  \right)\Pi_i \theta^i\ket{0;\pm}.
\label{ferzero}
\end{eqnarray}
The normalization factor $N^{(f)}(F)$ is given by
\begin{eqnarray}
  N^{(f)}(F)=
\left\{
  \begin{array}{ll}
1 & (\mbox{for the NS-NS sector}), \\
\left(\det (1+F)\right)^{\zeta(0)}& (\mbox{for the R-R sector}).
  \end{array}
\right.
\label{factorf}
\end{eqnarray}
This resultant form coincides with the one given in ref.\
\cite{CLNY2}.


\subsection{BRST invariance of the boundary state}

The annihilation of the BRST operator on the boundary state is
important for the string loop correction to the equation of motion in
superstring theory \cite{293}. In this subsection we show that the
generalized boundary state constructed in the previous section is
indeed BRST invariant. 

The BRST charge is composed of relevant two terms in addition to the
pure ghost sector as 
\begin{eqnarray}
&& \QB=    \QB^{(L)} + \QB^{(F)} 
+(\mbox{ghost sector}), \\
&&\mbox{where} \qquad 
\QB^{(L)}\equiv \sum_{\pm}\sum_n L_{-n}^{\alpha, b} c_{n},\quad 
\QB^{(F)}\equiv \sum_{\pm}\sum_n F_{-n}^{\alpha, b} \gamma_n,
\end{eqnarray}
and we see the vanishing of these two operators $\QB^{(L)}$ and
$\QB^{(F)}$ respectively. 
For the former term $\QB^{(L)}$, the following expression is found to
be useful:
\begin{eqnarray}
\QB^{(L)}
=\frac{1}{\sqrt{\pi}}\oint 
\left[
  i\Pi_{\bar{c}}  \left\{ \cdots \right\}
 + c
 \left\{
   2\pi P_M\p_\sigma X^M +\frac{i}{2}
    \left(\psi_+\p_\sigma\psi_+ + \psi_-\p_\sigma\psi_- \right)
 \right\}
\right].
\label{qbl}
\end{eqnarray}
The first term in the integrand disappears due to the ghost boundary
condition $\Pi_{\bar{c}}=0$. Next, the term $P_M\p_\sigma X^M$ is
evaluated using the boundary condition (\ref{p+FXB}) as 
\begin{eqnarray}
2\pi  P_M\p_\sigma X^M \bnew \hs{-5mm}&&= 
2\biggl[ -F_{\mu\nu}[X] \p_\sigma X^\nu  -\frac{i}{8} 
\p_\mu F_{\nu\rho}[X] (\psi_+ \pm i \psi_-)^\nu (\psi_+ \pm i
\psi_-)^\rho\biggr]\p_\sigma X^\mu \bnew\nn\\
&& =-\frac{i}{4} 
\p_\mu F_{\nu\rho}[X] (\psi_+ \pm i \psi_-)^\nu (\psi_+ \pm i
\psi_-)^\rho\p_\sigma X^\mu \bnew.
\label{fromPX}
\end{eqnarray}
For the evaluation of the rest term in eq.\ (\ref{qbl}), 
we use a new condition satisfied by $\bnew$,
\begin{eqnarray}
  \biggl[ \p_\sigma(\psi_+ \mp i \psi_-)^\mu - F^\mu_{\;\;\nu}[X]
\p_\sigma(\psi_+ \pm i \psi_-)^\nu 
- \p_\sigma X^\rho \p_\rho F^\mu_{\;\;\nu}[X]
(\psi_+ \pm i \psi_-)^\nu\biggr]
\bnew =0,
\end{eqnarray}
which is derived by differentiating the fermionic boundary condition
(\ref{p-FpB}) by $\sigma$. Then we obtain
\begin{eqnarray}
\lefteqn{\frac{i}{2} 
\biggl[\psi_+\p_\sigma\psi_+ + \psi_-\p_\sigma\psi_- \biggr]\bnew}
\nn\\
&& =
\frac{i}{2} 
\biggl[\Half(\psi_+ \mp i \psi_-)^\mu \p_\sigma 
(\psi_+ \pm i \psi_-)_\mu
+ \Half(\psi_+ \pm i \psi_-)^\mu \p_\sigma (\psi_+ \mp i \psi_-)_\mu
\biggr]\bnew\nn\\
&&=\frac{i}{4} 
\p_\mu F_{\nu\rho}[X] (\psi_+ \pm i \psi_-)^\nu (\psi_+ \pm i
\psi_-)^\rho\p_\sigma X^\mu \bnew,
\end{eqnarray}
which exactly cancels the term from $P_M\p_\sigma X^M$,
(\ref{fromPX}). Hence the operator $\QB^{(L)}$ vanishes on the
generalized boundary state (\ref{defwhole}). 

Another operator $\QB^{(F)}$ is expressed as 
\begin{eqnarray}
  \QB^{(F)} =\frac{1}{2\pi}
\oint d\sigma 
\left[
  (\pi P_M-\p_\sigma X_M)\psi^M_+ \gamma_+
+
  (\pi P_M+\p_\sigma X_M)\psi^M_- \gamma_-
\right].
\end{eqnarray}
Using the boundary conditions for the superghost \cite{293}
\begin{eqnarray}
  \gamma_+ = \mp i \gamma_-,
\end{eqnarray}
in addition to the boundary conditions (\ref{p+FXB}), (\ref{p-FpB}),
we obtain 
\begin{eqnarray}
   \QB^{(F)}\bnew = \frac{1}{2\pi}\oint \biggl[
-\frac{i}{8}\p_\mu F_{\nu\rho}
(\psi_+ \pm i \psi_-)^\mu
(\psi_+ \pm i \psi_-)^\nu(\psi_+ \pm i \psi_-)^\rho \gamma_+
\biggr]\bnew,
\end{eqnarray}
which vanishes due to the Bianchi identity.

Summing up all together, we see the BRST invariance of the generalized
boundary state $\bnew$. Since in the generalized boundary state full
configuration  of the gauge field coupled to the boundary is included, 
this degree of freedom of the gauge field is found to be actually one
of the collective coordinates of the BRST equation. However, any
configuration of the gauge field is not allowed, since this
boundary state contains divergence in general. For special
configuration of the gauge field, the boundary state is not divergent
and well-defined, as in the case of the bosonic string theory in ref.\
\cite{KH}. We will see this in the next section.

\section{Correction to the D-brane action}
\label{Div}

In this section, we calculate the corrections to the D-brane
action, using the supersymmetric generalized boundary state. The
procedure was already explored in our previous paper \cite{KH}, and
here we shall extend the calculation given there to the supersymmetric
case. We expand the field strength by derivatives, whose number is at
most two. Therefore we are considering the first correction to the
D-brane action, that is $O(\alpha' \partial^2)$, in the slowly-varying
field approximation.


\subsection{Divergence in the boundary state}

Before proceeding to the explicit evaluation of the correction to the
D-brane action, we explain how to calculate the relevant parts of the
generalized boundary state.

As seen in the definition of the operator in eqs.\ (\ref{U[A]}) and
(\ref{Rdef}), the newly defined boundary state (\ref{defwhole})
contains short distance divergence on the worldsheet. Therefore,
before calculating the correction to the D-brane action, we must
obtain well-defined boundary state with no divergence, through a
certain ``renormalization'' procedure. 

It would be hard to evaluate the operator $U[A]$ and $U[F]$ without
using the explicit functional form of the gauge field. Here we adopt
the Taylor expansion for the gauge field, and ignore the higher
derivative terms. We keep only the terms in which the number of the
derivatives acting on the gauge field strength is at most two. 
Then in this approximation, we can extract the short distance
divergence (or normal-ordering divergence) which exists in the
boundary state, with use of the simple linear boundary
conditions. Interestingly, the divergence in the generalized boundary
state can be perfectly absorbed into the redefinition of the gauge
field: 
\begin{eqnarray}
  A^{\rm red}_\mu(x) \equiv A_\mu(x) + \beta[A] \; \zeta(1).  
\label{eq:redefa}
\end{eqnarray}
The boundary state can be described by creation operators and this
redefined gauge field with finite coefficients. 
Of course, this redefinition includes the normal ordering
divergence $\zeta(1)$. For obtaining the well-defined boundary state,
we are forced to put the coefficient of the divergent part to be zero:
$\beta[A]=0$. This is the ``renormalization'' procedure for obtaining
the well-defined boundary state.

After putting this constraint on the gauge field, we can proceed to
the evaluation of the correction to the D-brane action. In the
following, after the explicit calculation, we will see that the
functional $\beta[A]$ in the above redefinition (\ref{eq:redefa}) is
actually identical to the familiar beta-function in the superstring
$\sigma$ model. Thus we interpret this constraint as the worldsheet
conformal invariance constraint. 

These facts which will be explained in detail below are just 
analogous to the situation of the generalized boundary state in the
bosonic string theory explored in ref. \cite{KH}. 
 

Now, let us present the detailed calculation. First, write explicitly
the zero mode part of the coordinate scalar and fermions as
\begin{eqnarray}
  \label{non-zero}
  X^\mu(\sigma) = x^\mu + \wt{X}^\mu(\sigma), \qquad
  \psi^\mu_\pm = \psi^\mu_{0\pm} + \wt{\psi}_{\pm}.
\end{eqnarray}
The zero mode of the fermion $\psi^\mu_{0\pm}$ exists only in the
Ramond sector. The gauge field depends arbitrarily on $X$ and we
Taylor-expand $A_\mu\left(X(\sigma)\right)$ around the zero mode 
$x^\mu$:
\begin{eqnarray}
A_\mu(X)=A_\mu(x) + \wtX^\nu\;\p_\nu A_\mu(x) +
\frac{1}{2}\wtX^\nu\wtX^\rho\;\p_\nu\p_\rho A_\mu(x)+\cdots,\\
F_{\mu\nu}(X)=F_{\mu\nu}(x) + \wtX^\rho\;\p_\rho F_{\mu\nu}(x) +
\frac{1}{2}\wtX^\rho \wtX^\delta \;\p_\rho \p_\delta F_{\mu\nu}(z)
+\cdots.
\end{eqnarray}
We decompose the unitary operator concerning the zero modes of $X$ as 
\begin{eqnarray}
U[A]=V U_0,\qquad U[F] = V^{(\rm f)} U_0^{(\rm f)}
\end{eqnarray}
where 
\begin{eqnarray}
&&U_0 = \exp\left(\frac{i}{2\pi}F_{\mu\nu}(x)
\oint\!d\sigma \;\p_\sigma \wtX^\mu \wtX^\nu 
\right),\\
&&U_0^{(\rm f)} = 
\exp
\left(
  \frac{1}{8\pi} F_{\mu\nu}(x)\oint d\sigma  
(\psi_+ \pm i \psi_-)^\mu (\psi_+ \pm i \psi_-)^\nu 
\right).
\end{eqnarray}
The rest part is written in an expanded form as
\begin{eqnarray}
\label{V}
&& V=
  1+{-i\over 3\pi}\p_\rho F_{\nu\mu}(x)\oint\! d\sigma 
    \, \p_\sigma \wtX^\mu\wtX^\nu\wtX^\rho
+{-i\over 8\pi}\p_\delta\p_\rho F_{\nu\mu}(x)
   \oint\! d\sigma 
    \, \p_\sigma \wtX^\mu\wtX^\nu\wtX^\rho\wtX^\delta
\nn\\
&&\hs{50mm}
   +\Half\left(
{-i\over 3\pi}\p_\rho F_{\nu\mu}(x)\oint\! d\sigma 
    \, \p_\sigma \wtX^\mu\wtX^\nu\wtX^\rho
\right)^2
+\cdots,\\
&&
V^{(\rm f)}=1+
  \frac{1}{8\pi} \p_\rho F_{\mu\nu}(x)\oint d\sigma  
(\psi_+ \pm i \psi_-)^\mu (\psi_+ \pm i \psi_-)^\nu \wtX^\rho
\nn\\
&&\hs{20mm} + 
  \frac{1}{8\pi} \p_\delta\p_\rho F_{\mu\nu}(x)\oint d\sigma  
(\psi_+ \pm i \psi_-)^\mu (\psi_+ \pm i \psi_-)^\nu 
\wtX^\rho \wtX^\delta
\nn\\
&&\hs{20mm} + \Half 
\left(
    \frac{1}{8\pi} \p_\rho F_{\mu\nu}(x)\oint d\sigma  
(\psi_+ \pm i \psi_-)^\mu (\psi_+ \pm i \psi_-)^\nu \wtX^\rho
\right)^2+\cdots.
\label{Vf}
\end{eqnarray}
In the above equations (\ref{V}) and (\ref{Vf}) we have kept 
terms in which the total number of the derivatives acting on
$F_{\mu\nu}(x)$ is at most two, as mentioned before. This is for the
leading order derivative correction to the D-brane action, of
$O(\alpha' \partial^2)$.  

Note that  $[U_0, V^{(\rm f)}]=0$ and that the zero mode contribution
in the unitary operator is summarized as 
\begin{equation}
U_0U_0^{(\rm f)}\ket{B(F\!=\!0)}=\bFx.
\label{UU}
\end{equation}
Here we have defined the boundary state for the constant field
strength by eqs.\ (\ref{bou}) and (\ref{expo}) as 
\begin{eqnarray}
  \ket{B(F)}\equiv \ket{B^{(b)}(F)} \otimes \ket{B^{(f)}(F)},
\end{eqnarray}
and the quantity $\bFx$ appearing in eq.\ (\ref{UU}) is obtained by
substituting $F(x)$ into the place of the constant $F$ in $\bF$.  
From these relations we obtain an expression
\begin{equation}
\bnew = V V^{(\rm f)}\bFx.
\label{VB}
\end{equation}
The base state $\bFx$ is written explicitly only by the creation
operators (see eq.\ (\ref{expo})), and satisfies the following
boundary conditions 
\begin{eqnarray}
\label{bc}
\left(\alpha^{(-)}_{n\mu} + \calO_\mu^{\;\nu}(x)
\alpha^{(+)}_{-n\;\nu}\right)\bFx=0,\\
\label{bcb}
\left(b_{r\mu}^{(-)}\pm i \calO_\mu^{\;\nu}(x)
b_{-r\nu}^{(+)}\right) \bFx=0,
\end{eqnarray}
which relate the annihilation operators on $\bFx$ to the creation
operators. Here $\calO(x)$ is an orthogonal matrix defined by 
\begin{eqnarray}
\calO_\mu^{\;\;\nu}(x) \equiv \left( {1-F(x) \over 1+F(x)}
\right)_\mu^{\;\nu}.
\end{eqnarray} 
In order to extract the normal ordering divergence in the
boundary state (\ref{VB}), we change all the annihilation operators in
the derivative perturbation $V$ into the creation ones with use of
these boundary conditions (\ref{bc}) and (\ref{bcb}).

We have accomplished this procedure for $V$ and $V^{(f)}$ to the
order mentioned above, and present here only the result (see the app.\ 
\ref{app:keisan} for the detailed calculation). The result is the same
as the case of the bosonic string theory \cite{KH}. The boundary state
to this order, expressed by only the creation operators, contains
$\zeta(0)$ and $\zeta(1)$ divergences. After regularizing one of the
divergences as $\zeta(0)=-1/2$, we are left with an intrinsic
short-distance divergence $\zeta(1)$. However, the $\zeta(1)$
divergent terms are arranged in a form in which the divergence is
perfectly absorbed into the redefinition of the field strength
$F(x)$ as  
\begin{eqnarray}
  \label{Bcreate}
  \bnew = \ket{B(F^{\rm red}(x))} 
+ 
\left[
\begin{array}{l}
\mbox{creation operators}\\
\mbox{with finite coefficients}
\end{array}
\right]
\bFx
\end{eqnarray}
where the redefined field strength $F^{\rm red}(x)$ is given by
\begin{eqnarray}
  \label{Fren}
\lefteqn{F^{\rm red}_{\mu\nu}(x) \equiv F_{\mu\nu}(x)
+ {1\over 2}\zeta(1)\left( {1\over 1\!\!+\!\!F}\right)^{\rho\delta}
\p_\rho\p_\delta F_{\mu\nu}}
\nn\\
 && {}\hs{15mm}+ {1\over 4}\zeta(1)
\left[  
\left\{
\left( {1\over 1\!\!+\!\!F}\right)^{\lambda\rho}
           \!\!\p_\nu F_{\rho\delta}
\left( {1\over 1\!\!+\!\!F}\right)^{\delta\kappa}\!\!
(\p_\lambda F_{\kappa\mu}+\p_\kappa F_{\lambda\mu})
\right\}
-
\left\{ \mu\leftrightarrow\nu\right\}
\right].
\end{eqnarray}
This result is precisely in the same form as in the bosonic string
case \cite{KH}. This redefinition of the gauge field strength is again
related to the redefinition of the gauge field as  
\begin{eqnarray}
\label{divA}
  A^{\rm red}_\mu(x) \equiv A_\mu(x)+\Half \zeta(1) \left(
1\over{1-F(x)^2}\right)^{\lambda\nu}\p_\lambda F_{\nu\mu}(x).
\end{eqnarray}
Hence for the well-defined boundary state which is not divergent, we
should put the coefficient of $\zeta(1)$ in the above redefinition to
zero. This is a constraint which the gauge field in the boundary state 
should satisfy. 

From the expression (\ref{divA}), one observes that the constraint on
the gauge field is identical to the vanishing of the $\beta$ function
in the superstring $\sigma$ model. Here we derived the constraint for
the background gauge field from the consistency of the generalized
boundary state. In general, string theory in some back ground field
put some restriction on the background when the consistency such as 
the conformal invariance is required. In our case, the
well-definedness of the boundary state has put such a constraint. 
Note that hereafter we adopt the background gauge field which
satisfies the constraint $\beta[A]=0$.  

The fact that the divergence (\ref{Fren}) takes the same form as in
the bosonic string theory has been expected from the string $\sigma$
model  calculation. This is because, in ref.\ \cite{BSPT} the
divergence in the superstring $\sigma$ model was shown to be equal to
the one in the bosonic string \cite{ACNY} within the one-loop
calculation. 

Similar to the bosonic string theory \cite{KH}, the divergence encoded 
in the field strength would be interpreted also from the superstring
$\sigma$ model loop calculation. The divergence in the boundary state
might correspond to the one-loop divergence against the propagator on
the boundary of the string worldsheet. (For details, see ref.\
\cite{KH}.)


\subsection{Correction to the D-brane action}
\label{D-brane}

Now we have obtained the well-defined boundary state with no
divergence, let us consider the coupling between the closed string
excitation and the boundary degrees of freedom. For obtaining  the
massless closed string coupling in the D-brane action, we extract the
relevant massless part of the boundary state. Using the boundary
state, we get information on the terms linear in the closed string
massless modes. 

Although the divergence immanent in the supersymmetric boundary state
is found to be equal to the one in the bosonic string case, the finite
part (the second term on the RHS of eq.\ (\ref{Bcreate})) has
different structures. In the boundary state, the part which is
relevant for the D-brane action is concerning the emission of the
massless states of the closed string. The projector which extracts the
relevant emission mode is studied recently in ref.\ \cite{fdp}. In our
language, the modes in $V$ relevant for the massless emission is only
the constant part (since if one excite $\alpha$ creation operator, the 
state becomes massive). In addition, the relevant part in $V^{(f)}$ is
the constant mode and $b^{(-)}_{-1/2}b^{(+)}_{-1/2}$ (which is
necessary only for the NS-NS sector).

It is shown in app.\ \ref{app:keisan} that all of these relevant modes
do not appear in the second term on the RHS of eq.\ (\ref{Bcreate})),
after changing all the oscillators in $V$ and $V^{(f)}$ into the
creation operators. Especially, the constant mode which exists in the
bosonic  string case, stemming from $V$, is canceled exactly by the
constant term coming from fermionic contribution $V^{(f)}$.

Therefore we conclude that, as for the part which can be
extracted from the boundary state (more precisely, the couplings
linear in the closed string massless modes),  there is {\it no
  $O(\alpha'\p^2)$ correction } to the D-brane action. This is
consistent with expectation from the 
other works on the correction to the D-brane action, refs.\ \cite{BBG}
and \cite{Tse}. In  these literature, it was shown that there is no
$O(\alpha'\p^2)$ correction in the other sectors (the (curvature)$^2$
sector, and the Born-Infeld sector with purely open string gauge
fields, respectively). This property of the absence of the
$O(\alpha'\p^2)$ terms in the D-brane action seems to be universal in 
superstring theory.


\section{Incorporation of scalar field and T-duality}
\label{sec:higgs}

In our previous paper \cite{KH} and in the former part of this paper, 
we have studied only the gauge field as a boundary degree of freedom. 
In addition to the gauge field, there exist scalar fields as another
massless excitation on the D-brane. This scalar field, usually treated 
on an equal footing with the gauge field, represents the deformation
of the D-brane. 

In this section, we incorporate this scalar field into the boundary
state. Similar to the gauge field, the scalar field takes an arbitrary 
configuration. The T-duality naturally relates this scalar field and
the gauge field on the D-brane, hence we use this perturbative
duality so as to check the consistency.

\subsection{Incorporation of the scalar field}
\label{sec:inc}

For simplicity, we concentrate on the bosonic string theory in sec.\
\ref{sec:inc} and \ref{sec:divphi}. The
generalization to the superstring theory is straightforward, and it
will be briefly mentioned later.

Since the scalar field parameterizes the deformation of the D-brane in 
the static gauge, let us define the boundary state with the scalar
field $\phi^i(X)$ as  
\begin{eqnarray}
  \ket{{\cal B}^{(b)}[\phi,A]}\equiv 
\wt{U}[\phi] U[A] \ket{B^{(b)}(F\!=\!0)},
\label{BT}
\end{eqnarray}
with a translation operator
\begin{eqnarray}
  \wt{U}[\phi]\equiv \exp
  \left(
    -i\oint \! d\sigma \; P_i \phi^i(X)
  \right).
\label{transT}
\end{eqnarray}
Note that the arguments of the scalar $\phi$ are $X^\mu$, the
tangential coordinates. When $\phi$ is linear in $X$ as $\phi^i(X) =
\theta^i_\mu X^\mu$, then  we reproduce the result of the tilted
D-brane case in ref.\ \cite{our}. 

The boundary conditions which the above boundary state satisfies are
as follows. For the transverse directions, the D-branes are now
expected to be deformed to a surface specified by the scalar, and
actually we have  
\begin{eqnarray}
  \left[
    X^i-\phi^i(X^\mu)
  \right]\ket{{\cal B}^{(b)}[\phi,A]}=0.
\label{bouTD}
\end{eqnarray}
On the other hand, the boundary conditions for the tangential
directions are also modified as
\begin{eqnarray}
  \left[
    \pi
    \left(P_\mu + P_i\p_\mu\phi^i(X)\right)
    +\p_\sigma X^\nu F_{\mu\nu}(X)
  \right]\ket{{\cal B}^{(b)}[\phi,A]}=0.
\label{bouTN}
\end{eqnarray}
This modification is natural in a sense that the combination $P_\mu +
P_i\p_\mu\phi^i(X)$ denotes the translation along the deformed surface 
of the D-brane. The ghost part is unchanged, as in ref.\ \cite{KH}.

Both the definition (\ref{BT}) and the boundary conditions
(\ref{bouTD}) and (\ref{bouTN}) can be understood
through the T-duality. This duality transformation is 
the exchange of the sign of the right-moving oscillators :
T-duality in the $M$-th direction is defined as $\alpha^{M(-)}
\rightarrow -\alpha^{M(-)}$. Under this transformation, $X$ is
exchanged for the conjugate momentum as
\begin{eqnarray}
  \p_\sigma X^M \leftrightarrow \p_\tau X^M (= -\pi P^M).
\label{Td}
\end{eqnarray}
Now, consider the situation with vanishing scalar field on the
D$p$-brane. Taking the T-duality (\ref{Td}) in one of the tangential
direction $\mu=p$ in the boundary condition (\ref{eigen}), then we
obtain new boundary conditions 
\begin{eqnarray}
  \pi P_p + \p_\sigma X^\nu F_{p\nu}(X)=0,\\
  \pi P_{\wt{\mu}} + \p_\sigma X^{\wt{\nu}} F_{\wt{\mu}\wt{\nu}}(X)
+\p_\sigma X^p F_{\wt{\mu}p}(X)=0,
\end{eqnarray}
on the T-dualized boundary state. The new tangential index
$\wt{\mu}$ runs from $0$ to $p-1$. Using a relation
\begin{eqnarray}
  F_{p\nu}=-\p_\nu A_p 
\end{eqnarray}
and defining new scalar field as $\phi_p \equiv -A_p$, then we
reproduce eq.\ (\ref{bouTN}) for D($p\!-\!1$)-brane and 
\begin{eqnarray}
  \p_\sigma (X^p - \phi^p) \ket{\cal B}=0.
\end{eqnarray}
This is consistent with the boundary condition in the transverse
direction (\ref{bouTD}), and only the zero mode part is not 
reproduced.\footnote{The
  discrepancy on the zero mode part is owing to the fact that we are
  considering only the oscillator part of the T-duality
  transformation, (\ref{Td}). If the target space is compactified, 
  winding modes appear and it becomes possible to treat zero modes
  simultaneously with oscillating modes. In this paper, uncompactified 
  flat spacetime is assumed.}
If one is not concerned with the zero modes, it is also possible to
understand directly the definition (\ref{BT}) as a 
result of the T-duality. Taking the T-dualities of the operator
$U[A]$, then we obtain the definition of the boundary state with the
scalar field (\ref{transT}).

In spite of the introduction of the scalar field, the boundary state
is still BRST invariant. In the case of the bosonic string theory, we
need to verify the vanishing of the quantity $P_M \p_\sigma X^M$
\cite{KH} for the BRST invariance. With use of the boundary  
conditions (\ref{bouTD}) and (\ref{bouTN}), this quantity is evaluated 
as 
\begin{eqnarray}
P_\mu \p_\sigma X^\mu + P_i \p_\sigma X^i
 = P_i (\p_\sigma X^i - \p_\sigma X^\mu \p_\mu \phi^i (X))
-\frac1\pi \p_\sigma X^\mu \p_\sigma X^\nu F_{\mu\nu}(X)=0.
\end{eqnarray}
This BRST invariance is seen clearly from the T-duality. Actually, The
operator $P_M \p_\sigma X^M$ (and the BRST charge $\QB$) is T-duality
invariant, hence if the previous boundary state with only the gauge
field is BRST invariant, then this is also the case for the one with
the scalar field. 

\subsection{Divergence in the boundary state}
\label{sec:divphi}

Using the T-duality nature of the definition of the boundary state
(\ref{BT}), as for the divergence in the boundary state, we are
trivially led to the same result as in the previous paper
\cite{our}. Since the translation operator $\wt{U}[\phi]$ is
transformed into the operator $U[A]$ by the T-duality, the procedure
of changing all the operators into the creation operators is actually
almost the same.  
All the divergences in the boundary state can be arranged in such a
way that they are absorbed into the redefinition of the gauge field
and the scalar field as
\begin{eqnarray}
  \label{Aren}
  A^{\rm red}_M(x) \equiv A_M(x)+\frac14 \zeta(1) 
J^{LN}(\theta, F)\p_L F_{NM}(x),
\end{eqnarray}
where $J(\theta, F)\equiv 1 + {\cal O}(\theta, F)$. The indices
$M,N,\cdots$ run through all the spacetime directions, and the gauge
fields with transverse index should be understood as scalar fields.
The matrix ${\cal O}(\theta, F)$ is defined in the same manner as
$\calO_M^{\;\;N}=(1-F)_N^{\;\;L}\{(1+F)^{-1}\}_{L}^{\;\;N}$, using the 
following definition:
\begin{eqnarray}
  F_{MN}\equiv
  \left(
    \begin{array}{cc}
      F_{\mu\nu} (x)& \theta^i_\mu (x)\\
      -\theta^j_\nu (x)& 0
    \end{array}
  \right),\qquad
{\rm where} \;\;\theta^i_\mu (x) \equiv \p_\mu \phi^i(x).
\label{fmn}
\end{eqnarray}
Explicitly for the scalar field, the redefinition becomes
\begin{eqnarray}
\label{redefp}
\phi^i_{\rm red} (x) \equiv \phi^i(x) + \frac14 \zeta(1)
J^{\mu\nu}(\theta, F)
\p_\mu \p_\nu \phi^i(x) + \cdots.
\end{eqnarray}
For the boundary state to be well-defined, these divergences should be 
eliminated, therefore the background configuration of the gauge field
and the scalar field should be restricted to the one which makes the 
coefficient of $\zeta(1)$ in eq.\ (\ref{Aren}) equal to zero.

The only difference in calculating the divergence is that in
$\wt{U}[\phi]$ there exists the zero mode $p^i$ of the operator $P^i$,
although in $U[A]$ the corresponding $\p_\sigma X^\mu$ has no zero
mode. This affects the zero mode of the transverse part of the
boundary state, that has been a delta function (see eq.\
(\ref{delta})).

Let us see concretely the zero mode part.
The relevant modes in the translation operator $\wt{U}[\phi]$ are
in the following:
\begin{eqnarray}
  \wt{U}[\phi] = \exp
  \left(-\phi^i(x)\frac{\p}{\p x^i}
    - \frac{1}{4\pi}\left[\oint \! d\sigma \;
  \wt{X}^\mu\wt{X}^\nu \right] 
   \p_\mu\p_\nu \phi^i(x)\frac{\p}{\p x^i} + \cdots 
  \right).
\label{wtuexp}
\end{eqnarray}
Here we expand the scalar field as 
\begin{eqnarray}
\phi^i(X) = \phi^i(x) + \wt{X}^\mu \p_\mu \phi^i(x) + 
\Half \wt{X}^\mu\wt{X}^\nu \p_\mu\p_\nu \phi^i(x) + \cdots.
\end{eqnarray}
Therefore the zero mode of the generalized boundary state is
\begin{eqnarray}
 \delta^{(9-p)}  \left( x^i - \phi^i(x) \right)
 - \frac{1}{4\pi}\left[\oint \! d\sigma \;
  \wt{X}^\mu\wt{X}^\nu \right] \p_\mu\p_\nu \phi^i(x)\frac{\p}{\p x^i}
\delta^{(9-p)}  \left( x^i - \phi^i(x) \right) 
+\cdots,
\label{zerod}
\end{eqnarray}
where the last omitted part consists of higher order
terms.\footnote{
Although the second term in eq.\ (\ref{zerod}) proportional to $\p\p
\phi$ corresponds to the $\p F$ mode in $U[A]$ by T-duality, this term 
contains already the derivative acting on the delta function, thus the 
number of the derivatives on the field strength ($\p\phi$) are
already two. Therefore, the term in proportion 
to $\p\p(\p\phi)$ or $\p(\p\phi) \p(\p\phi)$ are higher order terms. 
} The expression of 
the first term in the above representation (\ref{zerod}) is due to the 
first term in the exponent in eq.\ (\ref{wtuexp}). Evaluating the
divergence in the second term in eq.\ (\ref{zerod}) with use of
the boundary condition analogous to (\ref{bc}) now with the constant
field strength (\ref{fmn}), then we see that the result is 
\begin{eqnarray}
  -\frac{1}{4\pi} \cdot \pi\zeta(1) J^{\mu\nu}\cdot \p_\mu\p_\nu
  \phi^i(x) \frac{\p}{\p x^i}
\delta^{(9-p)}  \left( x^i - \phi^i(x) \right) 
\end{eqnarray}
This divergence can be absorbed into the delta function part in
the boundary state, with the redefinition (\ref{redefp}) as 
\begin{eqnarray}
  \delta^{(9-p)}  \left( x^i - \phi_{\rm red}^i(x) \right).
\end{eqnarray}

For general configurations of $A_\mu$ and $\phi^i$, the correction to
the D-brane action can be calculated using dimensional reduction
(with the identification $A_p = -\phi^p$), due to the
T-duality. However, as mentioned above, only the zero mode part of the
translation operator $\wt{U}[\phi]$ is different from T-dualized
$U[A]$, thus from this zero mode there appears a new finite part which
contributes to the D-brane action. The relevant term from the second
term in eq.\ (\ref{zerod}) is
\begin{eqnarray}
  \frac14 J^{\rho\lambda} J^{\mu\nu} \p_\nu\p_\rho \phi^i(x)
\frac{\p}{\p x^i}
\delta^{(9-p)}  \left( x^i - \phi^i(x) \right) 
\cdot\alpha^{(+)}_{-1 \lambda}\alpha^{(-)}_{-1 \mu}.
\label{corzero}
\end{eqnarray}
When contracting with the closed string massless modes, this 
contribution gives a new term in the correction to the bosonic D-brane
action, in addition to the contributions from the dimensional
reduction of the gauge field from ten-dimension.


\subsection{Supersymmetric case and correction to the D-brane action}

As for the supersymmetric boundary state studied in sec.\
\ref{sec:gene}, the scalar fields can be associated in the same
manner. Adopt the same redefinition $\alpha^{(-)}\rightarrow
-\alpha^{(-)}$ and $\phi^i(X)\equiv -A^i(X)$, and then for the
fermionic coordinates, define the T-duality transformation as 
\begin{eqnarray}
\psi_-\rightarrow -\psi_-.
\end{eqnarray}
All of the calculation of the divergence in the boundary state can be
read in correspondence with sec.\ \ref{Div} through the T-duality,
except for the zero modes of the operator $\wt{U}[\phi]$ considered in
the previous subsection. (Note that fermionic zero mode in the
R-R sector does not make mischief.) As for the divergence, this zero
mode contribution precisely gives the redefinition of the scalar field 
in the delta function, which results in no correction. Although in the 
bosonic string theory this zero mode contribution yields the
correction to the D-brane action (\ref{corzero}), in superstring case
this correction does not appear since the corresponding excitation is
massive.

As seen in sec.\ \ref{D-brane}, there is no $O(\alpha'\p^2)$
correction to the D-brane action in the superstring theory in the case 
of non-trivial gauge field configuration. Hence in the superstring
case, even if one incorporates the general brane deformation specified 
by $\phi(x^\mu)$, the D-brane action is not corrected to this order.


\section{Conclusion and discussion}

In this paper we have constructed generalized supersymmetric boundary
state which incorporates arbitrary configurations of massless fields
on the D-brane. In order to introduce such arbitrariness, we follow
ref.\ \cite{KH}, where the bosonic boundary state was 
generalized by a gauge transformation of a string field theory which
is an analogue of a string $\sigma$ model gauge transformation of
closed string theory.

The newly defined supersymmetric boundary state is BRST invariant, and
obeys the 
non-linear boundary conditions (\ref{p+FXB}) and (\ref{p-FpB}) which
are in the same form as derived in a superstring $\sigma$ model
\cite{BSPT}. Though this BRST invariance is verified for arbitrary
configuration of the background gauge field, the boundary state
contains short distance divergences originating in the products of
the coordinate fields on the worldsheet. For obtaining well-defined
boundary state, the divergent entry should be eliminated, so we
restrict the background gauge field configuration. This
constraint on the gauge field has been found to be identical to the
conformal invariance ($\beta(A)=0$) in a superstring $\sigma$ model
loop calculation. This has been checked  at least within the
next-to-leading order ($O(\alpha'\p^2)$ correction, corresponding to
the one-loop calculation in the $\sigma$ model). After this
subtraction, we have extracted finite corrections to the D-brane
action at this order, and have found that there exists no correction
of $O(\alpha'\p^2)$ to the coupling linear in the closed string
massless modes. For the R-R sector this property of no-correction is
expected from the fact that the coupling is relevant for the anomaly
cancellation. On the other hand, for the NS-NS sector the result is
non-trivial. The absence of the $O(\alpha'\p^2)$ correction to the
D-brane action seems to be universal (see refs.\ \cite{BBG,Tse,loop}).

The T-duality transformation relates Dirichlet and Neumann directions, 
hence it interchanges the gauge field with the scalar field on the
D-brane. This T-duality has enabled us to incorporate also the general
configuration of the scalar field into the boundary state. Since the
whole massless degrees of freedom are now incorporated in the boundary 
state formalism, it would be possible to analyze the non-trivial
configuration of the fields on the brane and various D-brane
configuration directly by the conformal field theory. 

Taking a BPS configuration which was previously studied as BIon
\cite{CM,Gibb,spike}, the divergence in the boundary state vanishes
and one obtains a boundary state for branes ending on another
brane. Especially adopting a BIon solution which represents a D-string 
stuck to a D3-brane, in the appendix B we have evaluated the boundary
state near the spike, and showed that in this region the form of the
boundary state approaches actually to the one of the D-string. This BIon 
boundary state are to be examined from various aspects in detail.

Another example for the application is the string junctions \cite{sj}
which have been already 
realized in two-dimensional gauge theory on D-strings \cite{Mukhi}.
Therefore it is possible to discuss the junctions with a single
boundary state. Or, using other BIon configurations, one can study
fundamental strings stuck to a D-brane. In this case, boundary states
representing fundamental strings\footnote{See ref.\ \cite{fdp} for a
  related discussion.} appear, and it is interesting to study how this
boundary states express the mechanism of the joining-splitting process
of the fundamental strings. 

On the other hand, most of intriguing brane configurations are related
to non-Abelian  
configurations of Yang-Mills-Higgs theory. One of the examples
belonging to this category is string junctions terminated on D3-branes 
($1/4$ BPS dyons in super Yang-Mills theory \cite{HHS}). A
non-Abelian version of the generalized boundary state would be
obtained with use of a path-ordered unitary operator
$U[A]$. Evaluation of this kind of operator seems to be difficult
technically. These remain to be studied in the future works.

\vspace{1.7cm}
\noindent
{\Large\bf Acknowledgments}\\[.2cm]
I would like to thank K.\ Furuuchi, A. Hashimoto, H.\ Hata,
N. Ishibashi, Y. Matsuo and S. Moriyama for valuable discussions and 
comments. This work is supported in part by Grant-in-Aid for
Scientific Research from Ministry of Education, Science, Sports and
Culture (\#3160).  I appreciate hospitality of the organizers of
Summer Institute `99 where a part of this work was discussed.

\vs{2cm}

\noindent
{\Large\bf Appendix}

\vs{-5mm}

\appendix

 
\section{Calculation of the divergence in the boundary state} 
\label{app:keisan}

In this appendix, we present the evaluation of the generalized
boundary state. As mentioned in sec.\ \ref{Div}, for extracting the
couplings between closed string excitations and boundary
degrees of freedom, the boundary state should be in a form where all
the oscillators are written only by creation operators. For
simplicity, we do not consider the scalar field in this appendix.

We calculate the divergences in the boundary state by changing
all the oscillators in $V$ and $V^{\rm (f)}$ (eqs.\ (\ref{V}) and
(\ref{Vf})) into creation operators explicitly, by using the boundary
conditions (\ref{bc}) and (\ref{bcb}) which hold on the boundary
state $\bFx$. We keep only terms up 
to $O(\alpha'\p^2)$ (at most two derivatives on the field strength
$F$), therefore the quantity to be evaluated is
\begin{eqnarray}
1 + (V-1) + (V^{\rm (f)}-1) + V_{\rm mix},
\label{V-1}
\end{eqnarray}
where the last term $V_{\rm mix}$ contains contributions from
both the bosonic part and the fermionic part,
\begin{eqnarray}
V_{\rm mix}\equiv\left[
\frac{-i}{3\pi}\p_\rho F_{\nu\mu}(x)\oint\! d\sigma \; 
\p_\sigma \wt{X}^\mu \wt{X}^\nu \wt{X}^\rho
\right]
\left[
  \frac{1}{8\pi} \p_a F_{bc}(x)\oint d\sigma  
(\psi_+ \pm i \psi_-)^b (\psi_+ \pm i \psi_-)^c \wtX^a
\right].
\end{eqnarray}
We shall evaluate these three terms $(V-1)$, $(V^{\rm (f)}-1)$, and 
$V_{\rm mix}$ respectively in the following. 
\begin{enumerate}
\item
For the bosonic part
$V-1$, we already have the result in ref.\ \cite{our}, since the
boundary condition for the bosonic oscillators in the supersymmetric
boundary state is the same as in the bosonic boundary state. The
result is as follows: First, the $\zeta(1)$ divergences in $V$ can be 
completely absorbed into the redefinition of the gauge field (or the
gauge field strength) of the bosonic part of the boundary state
$\bFx$. This redefinition is given by eq.\ (\ref{Fren}). Second, the
rest finite corrections to the boundary state is
\begin{eqnarray}
\; {1\over 64}
\p_\gamma F_{\alpha\beta}\p_\rho F_{\mu\nu}
\left[
J^{\rho\gamma}J^{\beta\nu}(J^{\mu\alpha}\!+\!J^{\alpha\mu})
\right]\bFx
+
\left[
\begin{array}{l}
\mbox{even number of}\\
\mbox{$\;\;\alpha$ oscillators}
\end{array}
\right]\bFx.
\label{corbos}
\end{eqnarray}
The second term in the above equation does not contribute to the
D-brane effective action\footnote{
In the boundary state, the terms which contribute to the D-brane
action is of the form
\begin{eqnarray}
 {\rm const.} \times \bFx 
\qquad {\rm or}   \qquad
b^{(+)\dagger}b^{(-)\dagger}\bFx.
\end{eqnarray}
Note that this is not the case for the bosonic string. See ref.\
\cite{our}.}, since these excitations are related 
to the coupling between the boundary and the massive excitations of
the closed string:  $(\alpha^\dagger)^2 +  (\alpha^\dagger)^4 +
(\alpha^\dagger)^6$. Hence the relevant correction is the first term
in eq.\ (\ref{corbos}).

\item
The last term $V_{\rm mix}$ in (\ref{V-1}) does not contribute both to
the divergence in the boundary state and to the D-brane effective
action. A possible contraction which may bring out the divergence is 
\begin{eqnarray}
  \wick{3}{
\oint\! d\sigma \; 
\p_\sigma \wt{X}^\mu \wt{X}^\nu <1{\wt{X}^\rho}
\oint d\sigma  
(\psi_+ \pm i \psi_-)^b (\psi_+ \pm i \psi_-)^c >1{\wtX^a}
}.\label{contmix}
\end{eqnarray}
(This is because  if one contract two of the three $X$'s in the first
integral, the rest single $X$ has no zero mode and therefore the whole
quantity vanishes. This is also the case for the fermion contraction.)
Evaluating the contraction (\ref{contmix}), one can easily see that
there is no divergence $\zeta(1)$. Additionally, there is no
contribution to the D-brane, since the massless excitation part of the 
boundary state does not stem from this $V_{\rm mix}$, even after
the contraction in the way (\ref{contmix}).

\item
Finally, for the fermionic part $(V^{\rm (f)}-1)$, we evaluate 
three non-trivial terms in  eq.\ (\ref{Vf}). 
\begin{enumerate}
\item
First, the term
proportional to $\p F$ (the first line in eq.\ (\ref{Vf})) has no
divergence  
and it is safely changed to the form written only by the creation
operators. This is because the contraction between two $\psi$'s makes 
no sence due to the fact that $\wtX$ has no zero mode. 
\item
Secondly, let us 
see that the term proportional to $\p\p F$ (the second line in eq.\
(\ref{Vf})) contains 
divergence. Changing all the $\alpha$ oscillators in two $\wtX$'s into
creation operators, then we have
\begin{eqnarray}
&&  \oint\! d\sigma (\psi_+ \pm i\psi_-)^\mu (\psi_+ \pm i\psi_-)^\nu 
\wtX^\rho \wtX^\delta\nn\\
&&\hs{10mm}=
\left[(\alpha^\dagger)^2
  +\frac14 \zeta(1)  
\left(
  J^{\rho\delta} + J^{\delta\rho}
\right)
\right]
\oint \! d\sigma  (\psi_+ \pm i\psi_-)^\mu (\psi_+ \pm i\psi_-)^\nu. 
\label{contppF}
\end{eqnarray}
This equality holds only on the boundary state $\bFx$. The rest
fermionic contraction is 
\begin{eqnarray}
\wick{2}{\oint\! d\sigma <a{(\psi)^\mu} >a{(\psi)^\nu}\wtX^\rho
\wtX^\delta},
\label{contpp}
\end{eqnarray}
however this does not give any divergence.  Therefore, one sees the
single divergent term in eq.\ (\ref{contppF}) from the $\p\p F$ part
in $(V^{\rm (f)}-1)$. Referring to the definition of the 
exponent $R$ of the boundary state, eq.\ (\ref{Rdef}), this $\zeta(1)$ 
divergence in (\ref{contppF}) can be absorbed into the redefinition of
the gauge field strength in the boundary state as
\begin{eqnarray}
F^{\rm red}_{\mu\nu}(x) \equiv F_{\mu\nu}(x)
+ {1\over 2}\zeta(1)\left( {1\over 1\!\!+\!\!F}\right)^{\rho\delta}
\p_\rho\p_\delta F_{\mu\nu}.
\end{eqnarray}
This is a part of the total redefinition of the gauge field
(\ref{Fren}). 
The finite part after the redefinition of the gauge field
strength consists of the $(\alpha^\dagger)^2$ term in eq.\
(\ref{contppF}) and (\ref{contpp}). Both of these terms contain two
$\alpha$'s, hence do not couple to the massless modes of the closed
superstring. Therefore, the $\p\p F$ part does not give any correction
to the D-brane action.

\item
The evaluation of the final term in $(V^{\rm (f)}-1)$,
\begin{eqnarray}
  \oint\! d\sigma 
(\wt\psi_+ \pm i \wt\psi_-)^\mu (\wt\psi_+ \pm i \wt\psi_-)^\nu 
\wtX^\rho \cdot  \oint\! d\sigma 
(\wt\psi_+ \pm i \wt\psi_-)^b (\wt\psi_+ \pm i \wt\psi_-)^c \wtX^a,
\label{pfpf}
\end{eqnarray}
is found to be rather complicated, as in the case of the bosonic
string \cite{our}. For simplicity, we consider only $\wt\psi$ which is 
non-zero modes of the fermionic operator $\psi$. (The zero modes of
$\psi$ exists only in the R-R sector, and they will be treated
separately later.) We change all the oscillators in this (\ref{pfpf})
into creation operators, using the boundary conditions (\ref{bc}) and
(\ref{bcb}). After some straight forward calculation, the result is
found as
\begin{eqnarray}
  (\ref{pfpf}) = c + (b^\dagger)^2 + (\mbox{other terms})
\label{cbo}
\end{eqnarray}
on the boundary state $\bFx$. Here the first constant $c$ term and the
second term quadratic in $b$ contain $\zeta(1)$ divergence as seen in
the following, and the rest ``other terms'' consisting of 
$(b^\dagger)^2(\alpha^\dagger)^2$, $(b^\dagger)^4$ and 
$(b^\dagger)^4(\alpha^\dagger)^2$ have finite
coefficients. Furthermore, these ``other terms'' couple to the massive
excitations of the closed superstring and thus have no contribution to
the D-brane action. The first constant term $c$ is given as follows:
\begin{eqnarray}
 (2\pi)^2
\Biggm[
\sum_{(1)}\frac1{4m}J^{\rho a}
  \left(
    J^{\nu c}J^{b \mu} - J^{\mu c} J^{b\nu}
  \right)
+
\sum_{(2)}\frac{-1}{4m}J^{a \rho}
  \left(
    J^{\nu c}J^{b \mu} - J^{\mu c} J^{b\nu}
  \right)
\nn\\
+
\sum_{(3)}\frac{-1}{4m}J^{a \rho}
  \left(
    -J^{\nu c}J^{\mu b} + J^{\mu c} J^{\nu b}
  \right)
+
\sum_{(4)}\frac{-1}{4m}J^{a \rho}
  \left(
    J^{c \nu} J^{\mu b} - J^{c \mu} J^{\nu b}
  \right)
\nn\\
+
\sum_{(5)}\frac{1}{4m}J^{\rho a}
  \left(
    J^{c \nu}J^{\mu b} - J^{c \mu} J^{\nu b}
  \right)
+
\sum_{(6)}\frac{1}{4m}J^{\rho a}
  \left(
    -J^{c \nu} J^{b \mu} - J^{c \mu} J^{b \nu}
  \right)
\Biggm].
\label{JJJ}
\end{eqnarray}
\begin{figure}[tdp]
\begin{center}
\leavevmode
\epsfxsize=5 cm
\epsfbox{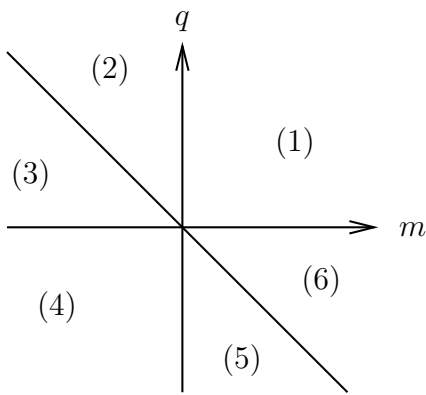}
\put(-40,92){(1)}
\put(-110,120){(2)}
\put(-140,80){(3)}
\put(-130,30){(4)}
\put(-60,10){(5)}
\put(-30,40){(6)}
\put(7,60){$m$}
\put(-78,140){$q$}
\caption[]{The region for summation in eq.\ (\ref{JJJ}).}
\label{fig:reg}
\end{center}
\end{figure}

The region for summation is depicted in fig.\ \ref{fig:reg}, where 
the index $q$ stems from the fermionic contraction. The summation of
$1/m$ in each region is given as 
\begin{eqnarray}
  \sum_{(1)} \frac1m =
\left\{ 
  \begin{array}{ll}
    0&(\mbox{NS-NS}),\\
    \zeta(1)\zeta(0)& (\mbox{R-R}),
  \end{array}
\right.
\quad
&
\displaystyle
  \sum_{(2)} \frac1m =
\left\{ 
  \begin{array}{ll}
    \zeta(0)&(\mbox{NS-NS}),\\
    -\zeta(1)\zeta(0)+\zeta(0)& (\mbox{R-R}),
  \end{array}
\right.
\nn\\
\sum_{(3)} \frac1m =
\left\{ 
  \begin{array}{ll}
    -\zeta(0)&(\mbox{NS-NS}),\\
    \zeta(1)-\zeta(0)& (\mbox{R-R}),
  \end{array}
\right.
&
\displaystyle  \sum_{(i+3)} \frac1m =   -\sum_{(i)} \frac1m,\;
{\rm for}\;\; i=1,2,3.
\label{sum}
\end{eqnarray}
The difference between the NS-NS sector and the R-R sector stems from
the fact that the lattice point to be summed in the region consists of 
integer lattice (the NS-NS sector) or half-odd integer lattice (the
R-R sector) along the direction of axis $q$. For the NS-NS sector,
there is no lattice point on the line $q+m=0$ in fig.\ \ref{fig:reg},
while for the R-R sector, lattice points on this line will give the
zero mode contribution which will be dealt with later. One can see
from eq.\ (\ref{sum}) that the $\zeta(1)$ divergence exists only in
the R-R sector, and finite terms (after regularizing as
$\zeta(0)=-1/2$) for the R-R sector and the NS-NS sector are the
same. Substituting the summation (\ref{sum}) into (\ref{JJJ}), 
the $\zeta(1)$ divergent term in $c$ in eq.\ (\ref{cbo}) is found as 
\begin{eqnarray}
  (2\pi)^2 \zeta(1) J^{\rho a} J^{b \mu}(J^{c \nu}- J^{\nu c})
\label{divconst}
\end{eqnarray}
which exists only in the R-R sector, and the finite term common in
both the R-R and NS-NS sector is 
\begin{eqnarray}
  (2\pi)^2   \Half J^{\rho a} J^{c \nu}(J^{b \mu} + J^{\mu b}).
\label{corcons}
\end{eqnarray}
Referring to eq.\ (\ref{factorf}), only in the R-R sector there is a
front factor in the boundary state for the fermionic sector (in the
case of constant field strength). Thus the divergence (\ref{divconst})
in front of the boundary state $\bFx$ can be absorbed into the
front normalization factor of the boundary state through redefinition
of the gauge field strength as  
\begin{eqnarray}
F^{\rm red}_{\mu\nu}(x) \equiv F_{\mu\nu}(x)
+ {1\over 16}\zeta(1)
\left[  
\left\{
J^{\lambda\rho}\p_\nu F_{\rho\delta} J^{\delta\kappa}
(\p_\lambda F_{\kappa\mu}+\p_\kappa F_{\lambda\mu})
\right\}
-
\left\{ \mu\leftrightarrow\nu\right\}
\right].
\label{Frenl}
\end{eqnarray}
This is the last term in the redefinition (\ref{Fren}). On the other
hand, the finite contribution (\ref{corcons}) results in a final form 
\begin{eqnarray}
  -{1\over 64}\;
\p_\gamma F_{\alpha\beta}\p_\rho F_{\mu\nu}\;
\left[
J^{\rho\gamma}J^{\beta\nu}(J^{\mu\alpha}+J^{\alpha\mu})
\right]\bFx.
  \label{corfermi}
\end{eqnarray}
This is common for both the NS-NS and R-R sector. 
Furthermore, one observes that this finite constant factor exactly
cancels the finite constant factor stemming from the bosonic part
(\ref{corbos}). 

Secondly, we check that the divergence in the $(b^\dagger)^2$ term in
eq.\ (\ref{cbo}) can be absorbed also by the redefinition
(\ref{Frenl}). The explicit expression of the $(b^\dagger)^2$ term in
eq.\ (\ref{cbo}) is 
\begin{eqnarray}
&&(2\pi)^2\Biggm[  \sum_{(1)}\frac1m J^{a\rho}
  \left(
    -J^{d\nu}J^{\mu c}J^{bf}\;b_{-q\; d}^{(-)}\;b_{-q\; f}^{(+)}
    -J^{c\mu}J^{\nu f}J^{db}\;b_{-m-q\; d}^{(-)}\;b_{-m-q\; f}^{(+)}
  \right)\nn\\
&&\hs{20mm}+ \sum_{(1)}\frac{-1}m J^{a\rho}
  \left(
    -J^{dc}J^{\mu f}J^{b\nu}\;b_{-q\; d}^{(-)}\;b_{-q\; f}^{(+)}
    -J^{cf}J^{\nu b}J^{d\mu}\;b_{-m-q\; d}^{(-)}\;b_{-m-q\; f}^{(+)}
  \right)\nn\\
&&\hs{20mm}+
\sum_{(3)}\frac{-1}m J^{a\rho}
  \left(
    J^{dc}J^{\nu b}J^{\mu f}\;b_{-q\; d}^{(-)}\;b_{-q\; f}^{(+)}
\right)
\nn\\
&&\hs{20mm}+
\sum_{(6)}\frac{1}m J^{a\rho}
  \left(
    J^{dc}J^{\nu b}J^{\mu f}\;b_{-m-q\; d}^{(-)}\;b_{-m-q\; f}^{(+)}
\right)\Biggm].
\label{JJJJbb}
\end{eqnarray}
In obtaining this expression, we have used the antisymmetric property
of the indices $(b\leftrightarrow c)$ and $(\mu \leftrightarrow \nu)$,
and change of the region for summation: (i)$\leftrightarrow$(i+3) 
with $(m,q)\leftrightarrow (-m,-q)$ simultaneously.
Let us extract the $\zeta(1)$ divergence in this expression
(\ref{JJJJbb}). For example, in the last term in (\ref{JJJJbb}), there
is a $\zeta(1)$ divergence in the summation $1/m$ in the region (6),
as seen if we fix the excitation number $-m-q$ of the oscillator $b$.
In this manner, we obtain the divergence in $(b^\dagger)^2$ in eq.\
(\ref{cbo}) as 
\begin{eqnarray}
2 (2\pi)^2 \zeta(1)
\left[
J^{d\mu}J^{\nu c}J^{bf}-J^{db}J^{\nu c}J^{\mu f} 
\right]
\sum_{q>0}\;b_{-q\; d}^{(-)}\;b_{-q\; f}^{(+)}.
\end{eqnarray}
It is easy to show that this divergence can be absorbed into the
exponent of the boundary state (\ref{expo}) by the
redefinition (\ref{Frenl}). 

The finite contribution to the D-brane action from (\ref{JJJJbb})
should be of the form $b_{-1/2}^{(-)}\;b_{-1/2}^{(+)}$ in the NS-NS
sector. Extracting that part, the coefficient of the term is turned
out to be zero. Thus there is no contribution to the D-brane action
from this  $(b^\dagger)^2$ term.

Finally, we consider the zero mode contribution in the term
(\ref{pfpf}) in the R-R sector. Written explicitly the coefficients,
the zero mode contribution is
\begin{eqnarray}
  4\times \Half
  \left(\frac{1}{8\pi}\right)^2
\p_\rho F_{\mu\nu} \p_a F_{bc}
  \oint\! d\sigma \;
\theta^\mu (\wt\psi_+ \pm i \wt\psi_-)^\nu 
\wtX^\rho \cdot  \oint\! d\sigma \;
\theta^b (\wt\psi_+ \pm i \wt\psi_-)^c \wtX^a.
\label{zeropfpf}
\end{eqnarray}
The front factor $4$ is due to the antisymmetric nature of the indices
of $F$. Changing all the oscillators in (\ref{zeropfpf}) with use of
the boundary conditions (\ref{bc}) and (\ref{bcb}), we obtain 
\begin{eqnarray}
  \frac{-1}{32}\zeta(1)
     \p_\rho F_{\mu\nu} \p_a F_{bc}\theta^\mu\theta^b
  \left(
   J^{\nu c} J^{a\rho} - J^{c\nu}J^{\rho a}
  \right) + (\alpha^\dagger)^2 + (b^\dagger)^2 +
  (\alpha^\dagger)^2(b^\dagger)^2.
\end{eqnarray}
The last three terms do not contain any divergence, and have no
contribution to the D-brane action. The first divergent term can be
absorbed by the redefinition of the field strength (\ref{Frenl}),
into the fermionic zero mode part of the boundary state
(\ref{ferzero}).

\end{enumerate}\end{enumerate}

Summing up all contributions, we conclude that all the divergences in
the boundary state can be absorbed into the redefinition of the gauge
field strength (\ref{Fren}), and the rest finite contributions
relevant for the D-brane action vanishes.


\section{Boundary state for brane ending on brane}

Well-defined configurations of the introduced gauge field and scalar
field are the ones in which the divergent part of the redefinition
(\ref{Fren}) (and (\ref{redefp})) vanishes. 
Therefore for studying the well-defined boundary state,
one must substitute to $A_\mu$ and $\phi^i$ the solutions of this
constraint which is identical with the leading-order equations of
motion derived from the open superstring $\sigma$ model approach. 

One of the interesting solution is the BIon configuration (or called
``spike soliton'') \cite{CM,Gibb}. This configuration is known to be a
solution of the equations of motion corrected to all order in
the derivative expansion \cite{Th}. Therefore, if one assumes that the 
divergence in the boundary state exactly coincides with the divergence 
in the string $\sigma$ model to all order, then this BIon
configuration is the most natural among nontrivial configurations of
$A_\mu$ and $\phi^i$, for a well-defined boundary state. Another
interesting respect concerning this solution is that this BIon
configuration represents F-strings (or D-strings) ending on a D-brane
\cite{CM,spike}. Hence adopting the BIon solution we obtain boundary
states representing branes ending on another D-brane.

Let us consider a specific example of a D-string stuck to a D3-brane. 
The corresponding BIon solution is the BPS configuration for a point
magnetic charge in the worldvolume theory on the D3-brane:
\begin{eqnarray}
  B_a = \p_a \phi^9 (X)\qquad {\rm with}\qquad \phi^9 = 1/r, 
\qquad r\equiv \sqrt{(X^1)^2+(X^2)^2+(X^3)^2},
\end{eqnarray}
where $B_a$ is the magnetic field with $a=1,2,3$. Let us evaluate this 
singular operator $1/r$ by $\alpha'$ expansion. Put the center of mass
of the string attached to this boundary state at $x^1=\epsilon$ and 
$x^2=x^3=0$. Then expand the magnetic field around the center of
mass as 
\begin{eqnarray}
  B_1=-\frac1{\epsilon^2}+ {\rm oscil.}, \quad
  B_2= 0 + {\rm oscil.}, \quad  B_3= 0 + {\rm oscil.}
\end{eqnarray}
We evaluate the effect of the constant mode of the magnetic field. The 
contribution of the center of mass becomes
\begin{eqnarray}
  F_{MN}= 
\left(  \begin{array}{cccc}
0&0&0&1/\epsilon^2\\
0&0&-1/\epsilon^2&0\\
0&1/\epsilon^2&0&0\\
-1/\epsilon^2&0&0&0
  \end{array}\right),
\end{eqnarray}
where the each column and row denote the direction along
$x^1,x^2,x^3,x^9$, respectively. Calculating the matrix ${\cal O}$
from this $F$ and taking the limit $\epsilon\rightarrow 0$ (in this
limit we are approaching to the D-string region, along the conjecture
in ref.\ \cite{CM}), then we have
\begin{eqnarray}
  {\cal O} \rightarrow {\rm diag} (-1,-1,-1,-1).
\end{eqnarray}
This means that the original Neumann directions $x^1,x^2,x^3$ change
their signs in front of the bilinear combination of the oscillators in
the exponent of the boundary state (see eq.\ (\ref{aoa})) and become
Dirichlet directions, while the 9-th (originally Dirichlet) direction
also changes the sign so as to become a Neumann type. This result does
not depend on where we expand the magnetic field, so long as we take
the limit of approaching to the singular point $X_a \sim 0$. Therefore
we have examined that the singular center of the BIon corresponds to a 
D-string which extends to the 9-th direction.

\newcommand{\J}[4]{{\sl #1} {\bf #2} (#3) #4}
\newcommand{\andJ}[3]{{\bf #1} (#2) #3}
\newcommand{\AP}{Ann.\ Phys.\ (N.Y.)}
\newcommand{\MPL}{Mod.\ Phys.\ Lett.}
\newcommand{\NP}{Nucl.\ Phys.}
\newcommand{\PL}{Phys.\ Lett.}
\newcommand{\PR}{Phys.\ Rev.}
\newcommand{\PRL}{Phys.\ Rev.\ Lett.}
\newcommand{\PTP}{Prog.\ Theor.\ Phys.}
\newcommand{\hep}[1]{{\tt hep-th/#1}}


\begin{thebibliography}{99}

\bibitem{293}
   C.\ G.\ Callan, C.\ Lovelace, C.\ R.\ Nappi and S.\ A.\ Yost,
   \J{\NP}{B293}{1987}{83}.

\bibitem{CLNY2} 
   C.\ G.\ Callan, C.\ Lovelace, C.\ R.\ Nappi and S.\ A.\ Yost,
   \J{\NP}{B308}{1988}{221}.

\bibitem{CLNY3} 
   C.\ G.\ Callan, C.\ Lovelace, C.\ R.\ Nappi and S.\ A.\ Yost,
   \J{\PL}{B206}{1988}{41}.

\bibitem{PC} 
   J.\ Polchinski and Y.\ Cai, \J{\NP}{B296}{1988}{91}.

\bibitem{Ishi}
   N.\ Ishibashi, \J{\MPL}{A4}{1989}{251}.
   
\bibitem{ML}
   M.\ Li, \J{\NP}{B460}{1996}{351}, \hep{9510161}.

\bibitem{Pol}
   J.\ Polchinski, \J{\PRL}{75}{1995}{4724}, \hep{9510017}.

\bibitem{fdp}
  P.\ Di Vecchia, M.\ Frau, A.\ Lerda and  A.\ Liccardo,
  \hep{9906214}.

\bibitem{GG}
  M.\ B.\ Green and M.\ Gutperle, \J{\NP}{B476}{1996}{484},
  \hep{9604091}.

\bibitem{DD}
   C.\ Schmidhuber, \J{\NP}{B467}{1996}{146}, \hep{9601003} ; \\
   A.\ A.\ Tseytlin, \J{\NP}{B469}{1996}{51}, \hep{9602064} ; \\
   D.\ P.\ Jatkar and S.\ K.\ Rama, \J{\PL}{B388}{1996}{283},
   \hep{9606009}.

\bibitem{BBG} C.\ P.\ Bachas, P.\ Bain and M.\ B.\ Green,
    \J{JHEP}{9905}{1999}{011}, \hep{9903210}.

\bibitem{Tse} 
   O.\ D.\ Andreef and A.\ A.\ Tseytlin, \J{\NP}{B311}{1988/89}{205} ; 
   \J{\MPL}{A3}{1988}{1349}.

\bibitem{Kal}
   N.\ Kaloper and K.\ Meissner,
   \J{\PR}{D56}{1997}{7940}, \hep{9705193}.

\bibitem{KH}
   K.\ Hashimoto, \hep{9909027}, to appear in \J{\PR}{D61}{2000}{}.


\bibitem{our} K.\ Hashimoto and H.\ Hata, \J{\PR}{D56}{1997}{5179}, 
   \hep{9704125}.

\bibitem{front}
  C.\ G.\ Callan and I.\ R.\ Klebanov, \J{\NP}{B465}{1996}{473}, 
  \hep{9511173}.

\bibitem{BSPT}
   E.\ Bergshoeff, E.\ Sezgin, C.\ N.\ Pope and P.\ K.\ Townsend, 
   \J{\PL}{B188}{1987}{70}.

\bibitem{ACNY} 
   A.\ Abouselsaood, C.\ G.\ Callan, C.\ R.\ Nappi and S.\ A.\ Yost, 
            \J{\NP}{B280(FS18)}{1987}{599}.

\bibitem{loop}
    M.\ Shmakova, \hep{9906239} ;\\
    A.\ De Giovanni, A.\ Santambrogio and D.\ Zanon, \hep{9907214}.

\bibitem{CM} 
    C.\ G.\ Callan and J.\ M.\ Maldacena, 
    \J{\NP}{B513}{1998}{198}, \hep{9708147}.

\bibitem{Gibb}
    G.\ W.\ Gibbons, 
    \J{\NP}{B514}{1998}{603}, {\tt hep-th/9709027}.  

\bibitem{spike}
    K.\ G.\ Savvidy and G.\ K.\ Savvidy, 
    \hep{9902023} ;\\
    K.\ Hashimoto, \J{JHEP}{9907}{1999}{016}, \hep{9905162} ;\\
    D.\ Kastor and J.\ Traschen, \hep{9906237}.

\bibitem{sj}
    J.\ H.\ Schwarz, \J{\sl Nucl.\ Phys.\ Proc.\ Suppl.\
    }{55B}{1997}{1}, \hep{9607201}.

\bibitem{Mukhi}
    K.\ Dasgupta and S.\ Mukhi, \J{\PL}{B423}{1998}{261},
    \hep{9711094} ;\\
    K.\ Hashimoto, \J{\PTP}{101}{1999}{1353}, \hep{9808185}.

\bibitem{HHS}
    O.\ Bergman, \J{\NP}{B525}{1998}{104}, \hep{9712211} ; \\
    K.\ Hashimoto, H.\ Hata and N.\ Sasakura,
    \J{\PL}{B431}{1998}{303}, \hep{} \\ \hspace*{10mm}{\tt 9803127} ;
    \J{\NP}{B535}{1998}{83},  
    \hep{9804164} ; \\
    T.\ Kawano and K.\ Okuyama, \J{\PL}{B432}{1998}{338},
    \hep{9804139} ; \\
    K.\ Lee and P.\ Yi, \J{\PR}{D58}{1998}{066005}, \hep{9804174}.

\bibitem{Th} 
    L. Thorlacius, 
    \J{\PRL}{80}{1998}{1588}, \hep{9710181}. 

\end{thebibliography}
\end{document}